\documentclass[preprint,authoryear,12pt]{elsarticle}%
\usepackage{graphicx}
\usepackage{epsfig}
\usepackage{amssymb}
\usepackage{amsthm}

\def\deg{\hbox{$^\circ$}}

\def\gsim{\;\rlap{\lower 2.5pt\hbox{$\sim$}}\raise 1.5pt\hbox{$>$}\;}
\def\lsim{\;\rlap{\lower 2.5pt\hbo time lag between the formation x{$\sim$}}\raise 1.5pt\hbox{$<$}\;}
\def\la{\mathrel{\hbox{\rlap{\hbox{\lower4pt\hbox{$\sim$}}}\hbox{$<$}}}}
\def\ga{\mathrel{\hbox{\rlap{\hbox{\lower4pt\hbox{$\sim$}}}\hbox{$>$}}}}

\def\arcmin{\hbox{$^\prime$}}
\def\arcsec{\hbox{$^{\prime\prime}$}}

\def\farcm{\hbox{$.\mkern-4mu^\prime$}}

\def\ubvri{\hbox{$U\!BV\!RI$}\,}            

\def\aj{AJ}
\def\apj{ApJ}
\def\aa{A\&A}
\def\mnras{MNRAS}
\def\pasp{PASP}
\journal{New Astronomy}
\begin{document}
\begin{frontmatter}
\title{Stellar contents and Star formation in the young cluster Stock 18}
\author[ari]{Himali Bhatt\corref{cor1}}
\ead{mshimali@gmail.com}
\author[ari]{Ram Sagar}
\ead{sagar@aries.res.in}
\address[ari]{Aryabhatta Research Institute of observational sciencES (ARIES), Manora Peak, Nainital, India 263 129}
\author[ari]{J. C. Pandey}
\ead{jeewan@aries.res.in}

\begin{abstract}
We have carried out deep (V$\sim$21 mag) \ubvri photometric study of the star cluster Stock 18.
These along with 
archival  Infrared data have been used to derive  
the basic cluster parameters and  also to study the star formation processes
in and around the cluster region.
The distance to the cluster is derived as 2.8$\pm$0.2 kpc while its age is estimated as $6.0\pm2.0$ Myr.
Present study indicates that interstellar reddening is normal in
the direction of the cluster. 
The mass function
slope is found to be -1.37$\pm$0.27 for the mass range  1$<M/M_\odot<$11.9.
There is no evidence found for the effect of mass segregation in main-sequence stars of the cluster.
A young stellar population with age between 
1-2 Myr have been found in and around the cluster region.
The presence of IRAS and AKARI sources with MSX intensity map also show 
the youth of the Sh2-170 region.
 
\end{abstract}

\begin{keyword}
young cluster, star formation, mass function, mass segregation individual : Stock 18 
\end{keyword}

\end{frontmatter}

\section{Introduction} \label{sec:st_int}
Stock 18 is a young open cluster and situated in the vicinity of
the Sharpless region 170 (Sh2-170), which is almost circular H II region 
in Cassiopia with angular diameter of about $\rm{18\arcmin}$ ($\it{l}=$$\rm{117.62^{\deg}}$, $\it{b}=$$\rm{ +2.27^{\deg}}$; 
Roger et al. 2004). The nebula is excited by a centrally located 
main-sequence (MS) O9-type star, BD+63 2093p (Russeil, Adami \& Georgelin 2007) which is located at a  photometric distance of 2990 pc (Mayer \& Mac$\acute{a}$k 1973).
However, determination of cluster parameters as well as its photometric study are lacking,  which are very important to understand the association between the massive star, BD+63 2093p, and the cluster Stock 18.  In this paper, a comprehensive exploration of multi-wavelength data is presented to understand the global scenario of star formation in the Stock 18 region.  Deep optical (V$\sim$21 mag) 
$UBVRI$ photometric data for  $13\arcmin \times 13\arcmin$ region about the cluster Stock 18 is being reported for the first time in the present study.  
These along with the multiwavelength data are described in the next section while the cluster parameters are determined in the section 3. The results obtained from NIR data are discussed in section 4 while last section summarizes our results.

\section{Observations and data reduction} \label{sec:st_reduc}

\subsection{Optical data}  \label{sec:st_optdata}
\begin{figure*}
  \centering
  \includegraphics[width=12cm]{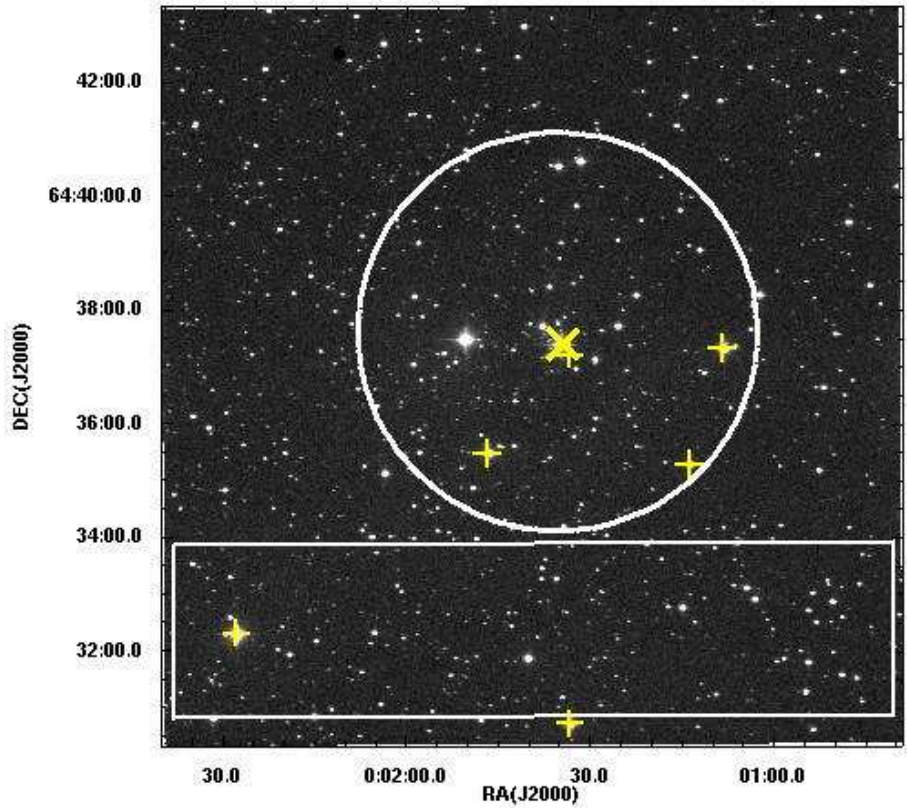}
  \caption{Observed image of Stock 18 from ST Nainital. The observed cluster and field regions are marked as circle and rectangle, respectively. Position of the star BD+63 2093p and other massive stars from Skiff (2009) are marked by the symbols of cross and plus, respectively.}
  \label{fig:st_image}
  \end{figure*}

The CCD \ubvri  photometric data were acquired 
using 104-cm Sampurnanand Telescope (ST) of 
ARIES, Nainital.
 The broad-band $UBVRI$ observations were standardized
by observing stars in the SA 98 field (Landolt 1992).
In Figure ~\ref{fig:st_image}  the observed cluster and field regions (see \S\ref{sec:st_rdp}) are marked by circle and rectangle, respectively.  The massive stars identified by  Skiff (2009) along with BD+63 2092p are also  marked 
in Figure~\ref{fig:st_image}.
The log of the
observations is given in Table~\ref{tab:st18_optlog}.

    The CCD data frames were reduced using computing facilities available at ARIES,
Nainital. Initial processing of data frames were done using {\sc Iraf} and {\sc Eso-Midas}
data reduction packages. Photometry of cleaned frames was carried out using {\sc Daophot II}
software (Stetson 1987). Calibration of the instrumental magnitude to the standard
system was done by using procedures outlined by Stetson (1992).

    For translating the instrumental magnitude to the standard magnitude, the 
calibration equations are derived  linear least-square fitting. These equations are given below:

\noindent
$u = U + (6.941 \pm 0.007) - (0.011 \pm 0.006)(U - B) + (0.516 \pm 0.012)X$\\
$b = B + (4.749 \pm 0.007) - (0.044 \pm 0.005)(B - V) + (0.274 \pm 0.009)X$\\
$v = V + (4.313 \pm 0.007) - (0.039 \pm 0.005)(V - I) + (0.157 \pm 0.009)X$\\
$r = R + (4.222 \pm 0.006) - (0.054 \pm 0.007)(V - R) + (0.104 \pm 0.007)X$\\
$i = I + (4.711 \pm 0.006) - (0.057 \pm 0.004)(V - I) + (0.047 \pm 0.007)X$\\

\noindent
    where $U, B, V, R$ and $I$ are the standard magnitudes, $u, b, v, r$ and $i$ are the instrumental aperture magnitudes normalized for 1 second of exposure time, and $X$ is the airmass. 
The standard deviations of the differences between 
 the magnitudes calibrated using above transformation equations 
and the standards given by Landolt (1992) are 0.035, 0.026, 0.025, 0.026 and 0.027 for 
U, B, V, R and I magnitudes, respectively.
The typical {\sc Daophot} errors in magnitude are found to be  large ($>$0.1 mag) 
for stars fainter than $V\approx$21 mag,
so the measurements beyond this magnitude are not considered.
At $V$ band, we 
could detect 2261 stars in nearly $13\arcmin\times13\arcmin$ observed region and their photometric 
magnitudes are given in Table~\ref{tab:datasample}\footnote{Full version of the Table is 
available only in electronic form.}.

For further study, it is very important 
to take into account the incompleteness
that may occur for various reasons (e.g. crowding of the stars). We used the {\sc Addstar}
routine of {\sc Daophot II} to determine the completeness factor (CF). The procedures
have been outlined in detail in earlier works (Sagar \& Richtler 1991, Pandey et al.
2001, 2005). The CF values are obtained to be 100\% for brighter stars (V$<$16 mag).
As expected, it decreases with faintness of the star and  
found to be nearly 80\% at V$\approx$ 21 mag for both 
 cluster region as well as field regions (see \S\ref{sec:st_rdp}).

\subsection{Archival data sets} \label{sec:st_otherset}

We have also used near-infrared (NIR),  mid-infrared (MIR) and far-infrared (FIR)
archival datasets from the surveys
such as Two micron all sky survey (2MASS), Infrared astronomical satellite (IRAS), Midcourse space experiment (MSX)
and a Japanese infrared space mission AKARI in the present study.
FIR data is essential to obtain an unbiased view of star formation and in particular to establish 
whether very young protostars are present (Zavagno et al. 2010), 
 especially those undetected at shorter wavelengths.
They represent the youth of the star forming regions and provide insight into the history of star formation.
Three IRAS point
sources namely IRAS 00001+6417, IRAS 23586+6412 and IRAS 23589+6421, 
and ten AKARI sources namely 
0001312+643829 ,
0001577+643734 ,
0001275+644107 ,
0001039+643907 ,
0001133+643135 ,
0000424+644218 ,
0002409+643405 ,
0000523+643026 ,
0001174+644607 and
0002530+644309
 have  been identified 
in the direction of cluster.
These sources have been detected at least in three bands.
AKARI and IRAS sources are different, however IRAS source IRAS00001+6417 is found within search radius $3\arcsec$ 
of the position of AKARI source 0002409+643405 and therefore, they may represent the same source.
 The presence of these infrared sources shows that the surroundings of massive star BD+63 2093p provide 
sites of ongoing star formation which is 
useful to study the star formation processes.

\section{Cluster Parameters}

\subsection{Radial density profile} \label{sec:st_rdp}

The cluster center is estimated iteratively by calculating average X and Y 
positions of stars with $V\leq18.0$ mag within 80 pixels from eye estimated 
center, until it converged to a constant value. The coordinates of the 
cluster center are found to be $\rm{RA_{J2000}= 00^h 01^m 35^s}$ and
 $\rm{DEC_{J2000} = 64^{\deg} 37{\arcmin} 37{\arcsec} }$. Using this method, a typical error
expected in locating the center is 5\arcsec. The center of the cluster coincides 
(within $\sigma$ limits) with the position of 
the massive star BD+63 2093p, as given by Mayer \& Mac$\acute{a}$k (1973). Therefore,
the massive star, BD+63 2093p is found to be exactly at the center of the cluster. 

\begin{figure*}
  \centering
  \includegraphics[width=12cm]{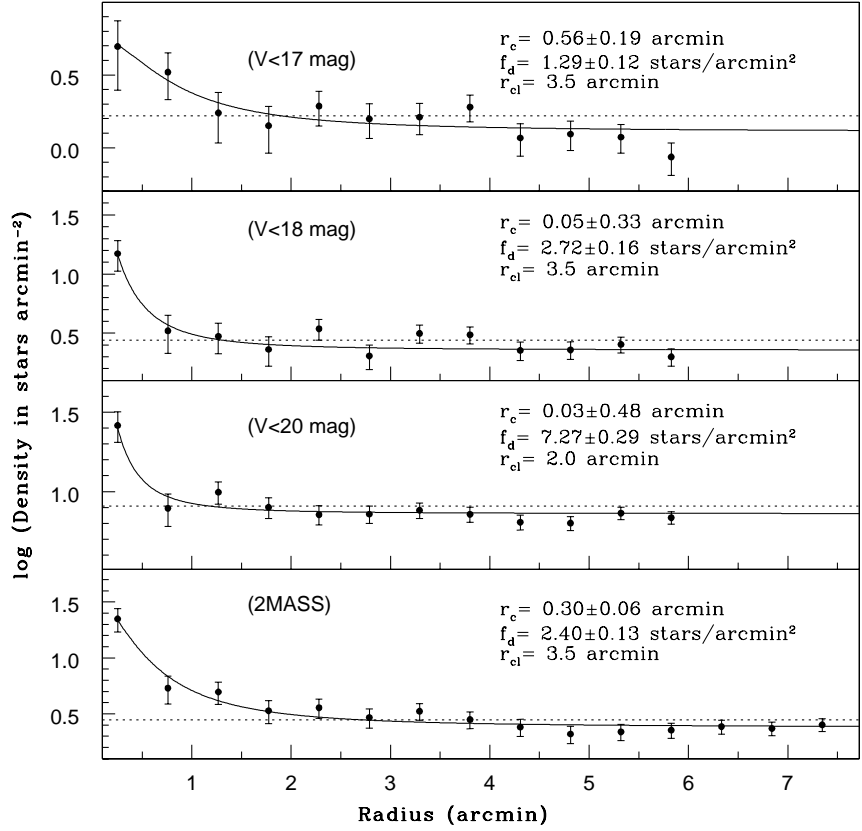}
  \caption{Projected radial stellar density profile of Stock 18.
    Dashed lines represent 3$\sigma$ levels above the field star density 
    and solid curve shows best fit to the empirical model of King (1962)
    .}
  \label{fig:st_rdp}
  \end{figure*}

     To determine the radial surface density, $\rho(r)$, we divided the
cluster into a number of concentric circles. Projected radial
stellar density in each concentric annulus was obtained by
dividing the number of stars in each annulus by its area and
the same are plotted in  Figure~\ref{fig:st_rdp} for various magnitude levels.
The error bars are derived assuming that the number of stars
in a concentric annulus follow the Poisson statistics. 

The $\rho(r)$ is parameterized as given by King (1962) 

  \begin{equation}
    \rho(r) \propto \frac { f_{\rm 0}} { 1+(r/r_{\rm c})^2}  
  \label{eqn:rdp}
  \end{equation} 

\noindent

where $r_{\rm c}$  and $f_{\rm 0}$ are the core radius of the cluster and
central star density, respectively. 
The field density, $f_{\rm b}$,  is derived using outer region 
($r>5\farcm6$) of the cluster and 3$\sigma$ of the field density is
 shown by dashed line in Figure~\ref{fig:st_rdp}. 
  We fit the function given by King (1962) to the
observed radial density distribution of stars by 
 Levenberg-Marquardt non-linear fitting routine 
and plotted by solid lines in  Figure~\ref{fig:st_rdp}.
The best-fit parameters are also given in Figure~\ref{fig:st_rdp}.
The core radius, $r_{\rm c}$, is obtained for various  magnitude levels 
(17$<$V$<$20), however errors are large in the magnitude range 18 and 20.
The cluster radius ($r_{\rm cl}$) is thus estimated to be 3\farcm50$\pm$0\farcm25, after considering
the radius at which cluster density is found to be 3$\sigma$ level above the field star 
density, marked by dotted lines in Figure~\ref{fig:st_rdp}. 
The estimated angular diameter, i.e. 5.7 pc, of the cluster is therefore
nearly three times smaller
than that of Sharpless region Sh-170 as given by Roger et al. (2004).

\subsection{Interstellar extinction} \label{sec:st_ext}

\begin{figure*}
  \includegraphics[width=12cm]{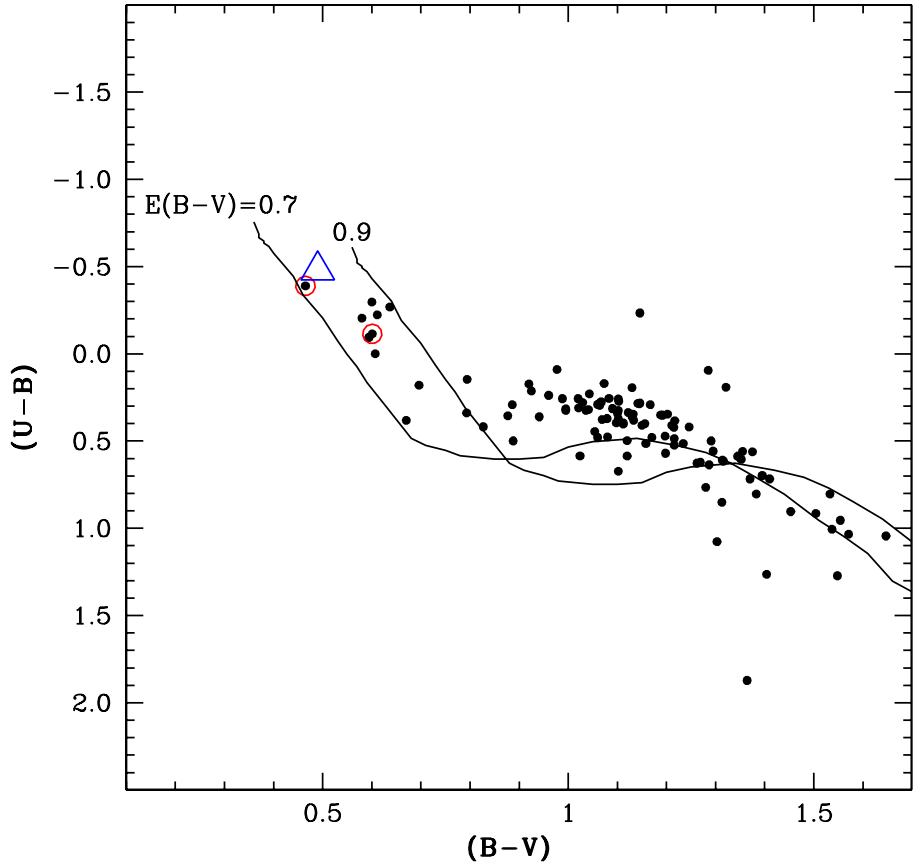}
  \caption{The $(B-V),(U-B)$ CC diagram for stars in the cluster 
    region.  Solid lines 
    are intrinsic MS reddened along the reddening line with 
    $E(B-V)=0.7$ mag and 0.9 mag.  BD+63 2093p is marked by triangle, and 
early B-type stars Sh 2-170 3 and LS I +64 9 present in our photometry are marked by open circles.}
  \label{fig:st_ccd}
  \end{figure*}

The interstellar extinction and the ratio of total-to-selective extinction $R_V = A_V /E(B -V)$
towards the cluster are important quantities  for accurate determination of the photometeric distance of the cluster.
The extinction towards the cluster region is estimated using the $(B-V), (U-B)$
color-color (CC) diagram shown in Figure~\ref{fig:st_ccd}, where the MS (Schmidt-Kaler 1982) is
shifted along the reddening vector having an adopted slope of $E(U - B)/E(B - V )$
= 0.72 to match the observations. It shows a spread in the observed sequence E(B - V ) = 0.7 to 0.9 mag
indicating the presence of differential reddening which is an indication of the youth of
the cluster.

 To study the nature of the
extinction law in the cluster region, we used $(V -\lambda )$ vs. $(B - V )$ diagrams,
 where $\lambda$ is one of the wavelengths of
the broad-band filters (R, I, J, H, K). This provides an effective method for separating the
influence of the normal extinction produced by the diffuse interstellar medium from
that of the abnormal extinction arising within regions having a peculiar distribution of
dust sizes (cf. Chini \& Wargau 1990, Pandey et al. 2000).
The slope of the distributions $m_{cluster}$  
for colors $(V-I)$, $(V-J)$, $(V-H)$ and $(V-K)$ 
with respect to $(B-V)$ color are 1.11$\pm$0.04, 2.08$\pm$0.07, 2.48$\pm$0.09 and 2.50$\pm$0.10, respectively.
These values are similar to that of normal reddening towards the direction of cluster.

The nature of interstellar reddening towards the cluster 
direction have also been analyzed using the color excess ratio method as described
by Johnson (1968). The intrinsic colors of the stars with spectral types earlier than A0
($\rm{V<14.4}$ mag) have been obtained using Q-method (Johnson \& Morgan 1953) 
and
iteratively estimated the reddening free parameter 
Q [$=(U-B)-X(B-V)$, where $X=E(U-B)/E(B-V)$], 
till the color excess ratios become constant within the photometric errors (cf. Joshi et al. 2008).
The color excesses are determined 
using color relation given by Caldwell et al. (1993) for $(U-B)$, $(B-V)$, $(V-R)$ and $(V-I)$ colors 
and by Koornneef (1983) for $(V-J)$, $(V-H)$ and $(V-K)$ colors. 
The mean values of color excess ratios $\frac{E(U-B)}{E(B-V)}$, 
$\frac{E(V-R)}{E(B-V)}$,
$\frac{E(V-I)}{E(B-V)}$,
$\frac{E(U-B)}{E(V-J)}$,
$\frac{E(B-V)}{E(V-J)}$,
$\frac{E(V-R)}{E(V-J)}$,
$\frac{E(V-I)}{E(V-J)}$,
$\frac{E(V-H)}{E(V-J)}$ and
$\frac{E(V-K)}{E(V-J)}$
are
0.77$\pm$0.06,
0.54$\pm$0.03,
1.28$\pm$0.09,
0.37$\pm$0.04,
0.49$\pm$0.04,
0.26$\pm$0.02,
0.62$\pm$0.02,
1.21$\pm$0.03 and
1.26$\pm$0.04, respectively.
All color excess ratios are in 
agreement within  $1 \sigma$ limit that expected for normal interstellar 
matter (Cardelli, Clayton \& Mathis 1989).
The ratio of total-to-selective extinction is estimated by using the 
relation ${R_V = 1.1E(V-K)/E(B-V)}$ (Whittet \& van Breda 1980) and it is found to 
be $2.9\pm0.3$, which is in agreement with the normal value of 3.1 within $1 \sigma$ limit, thus, indicating
 a normal grain size
in the direction of cluster.

\subsection{Optical colour magnitude diagram (CMD)} \label{sec:st_cmd}

\begin{figure*}
  \includegraphics[width=16.0cm]{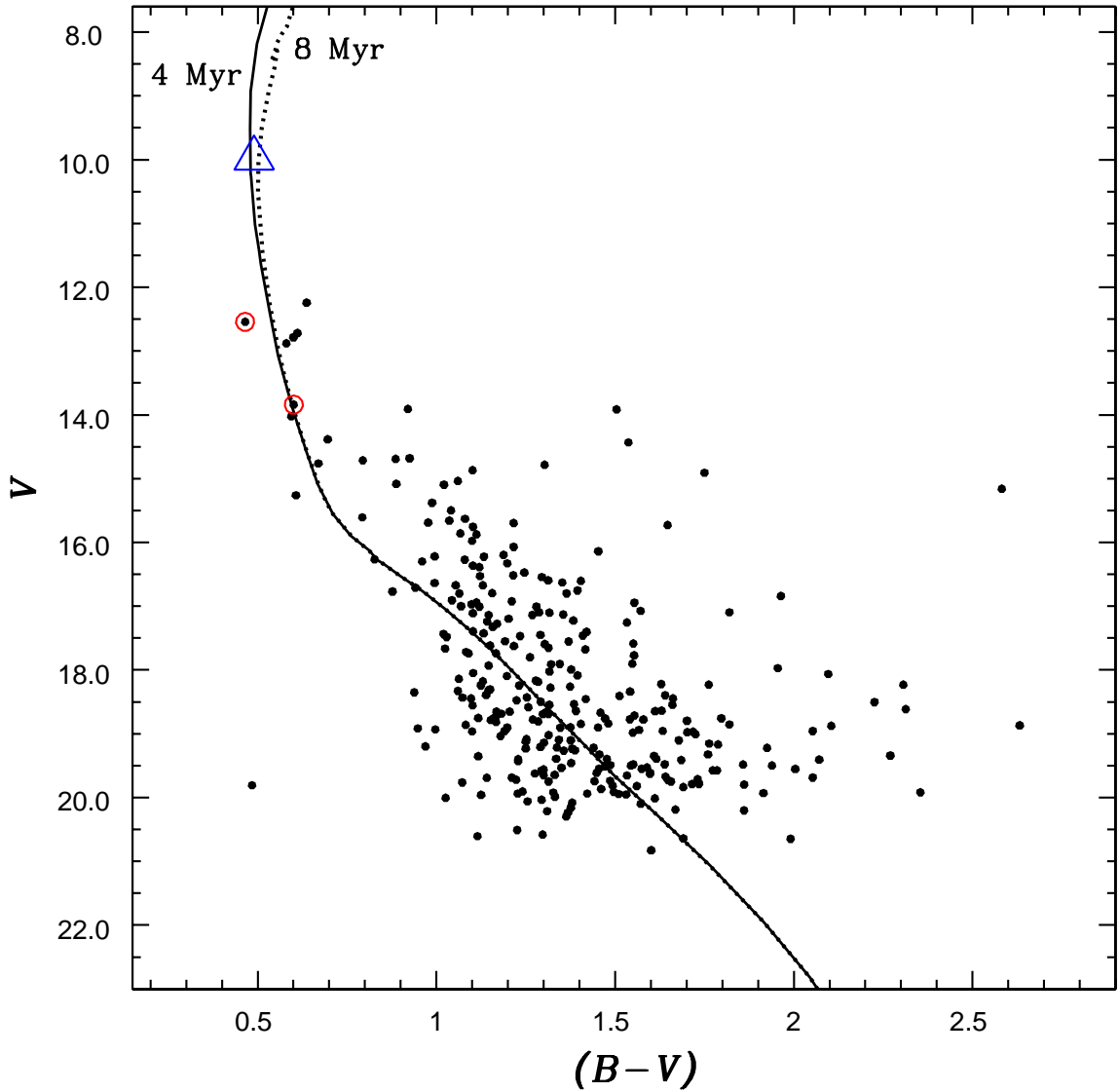}
  \caption{$V,(B-V)$ CMD for the cluster region.
    Theoretical isochrones from Girardi et al. (2002) are 
    shown with continuous solid and dotted lines for log(age)= 6.6 and 6.9 Myr, respectively.
    Early B-type stars Sh 2-170 3 and LS I +64 9 are present in our photometry and marked by open circles. 
    Massive star BD+63 2093p is marked by triangle and its $UBV$ data is taken from Mayer \& Mac$\acute{a}$k (1973). }
  \label{fig:st_cmdbv}
  \end{figure*}

The  $V,(B-V)$ CMD
 for the stars  in cluster region  is shown in Figure~\ref{fig:st_cmdbv}. The brightest star BD+63 2093p with spectral type O9 is marked by triangle while other two early B-type stars Sh2-170 3 and LS I+64 9 present in our photometry are marked by open circles.
The $UBV$ magnitudes of the star  BD+63 2093p  are taken from Mayer \& Mac$\acute{a}$k (1973), 
as it gets saturated even on our shortest exposure.
Using the minimum value of  $E(B - V )$= 0.70 mag and following relations $E(U - B)/E(B - V )$=0.72 and
$A_V$ = 3.1 $\times$ $E(B - V )$,
we visually fit the theoretical isochrones   
(Girardi et al. 2002), for log(age) = 6.6 (4 Myr) and 6.9 (8 Myr)
with  Z = 0.19 to the blue envelope of the observed MS.
Best-fit isochrone yields a distance modulus of $(m-M)_V$ = 14.44$\pm$0.15
may  corresponding to a distance of 2800$\pm$200 pc and age=6$\pm$2 Myr.
This distance estimation is in agreement with
the distance to the star BD+63 2093p (Mayer \& Mac$\acute{a}$k 1973).

\begin{figure*}
  \includegraphics[width=16.0cm]{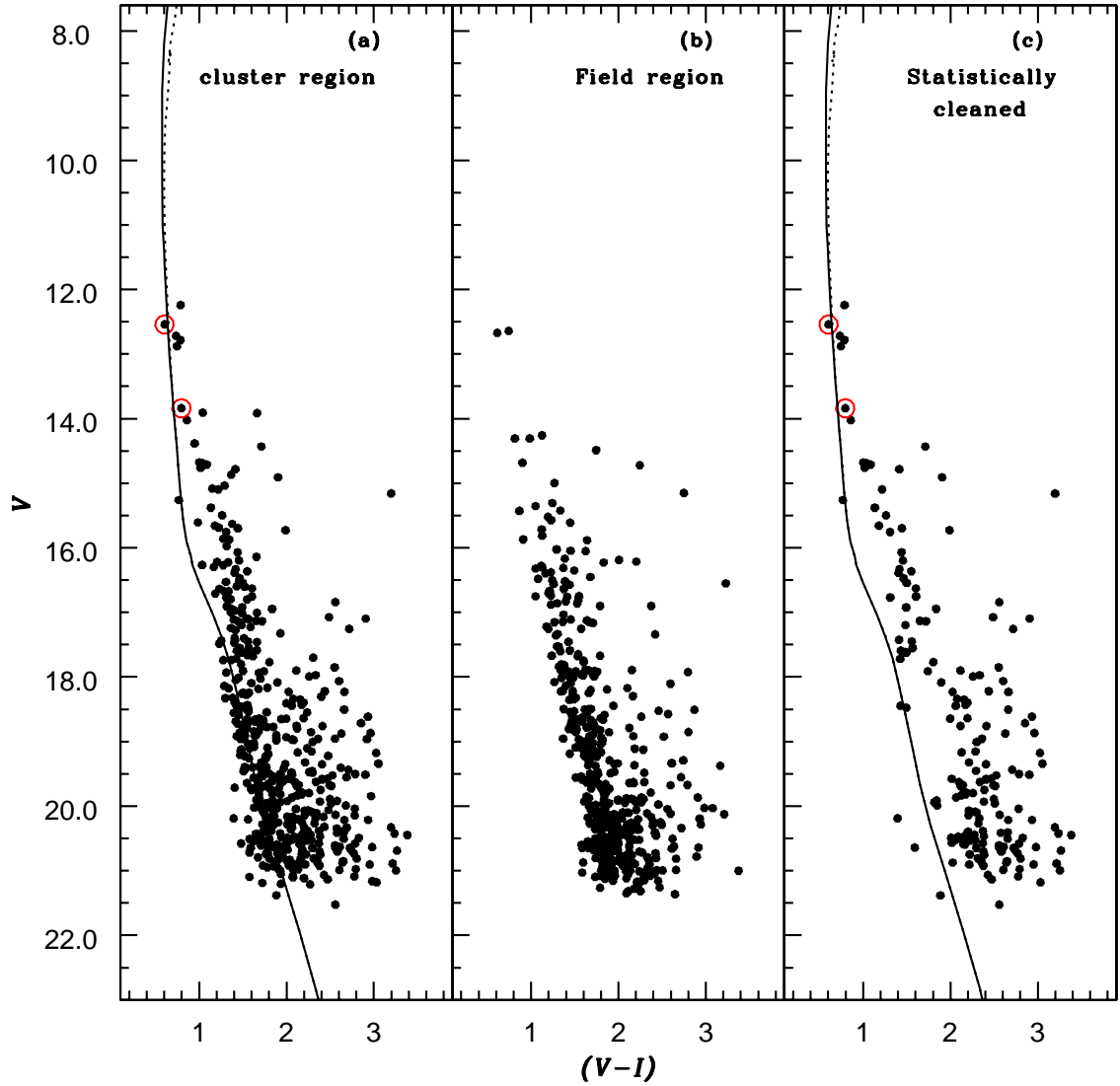}
  \caption{$V,(V-I)$ CMDs for (a) cluster region (b) field region (c) the statistically cleaned CMD for cluster.
    Theoretical isochrones from Girardi et al. (2002) are 
    shown with continuous  and dotted lines for log(age)= 6.6 and 6.9 Myr, respectively.
    Early B-type stars Sh 2-170 3 and LS I +64 9 are marked by open circles.}
  \label{fig:st_cmdvi}
  \end{figure*}

Due to better quantum efficiency of CCD detector in the red region, $V,(V-I)$ CMD is about a magnitude deeper in comparison to $V,(B-V)$ CMD. Hence,
$V,(V-I)$ CMD is useful to probe the stars at lower mass range and the same has done below.
Figures~\ref{fig:st_cmdvi} (a) and (b)  show $V,(V-I)$ CMDs for the cluster region and field region, respectively. The contamination due to field population is clearly visible in the CMD of cluster region.  To remove contamination of field stars from the MS  and PMS    sample, we statistically
subtracted the contribution of field stars from the CMD of the cluster region using the
procedure called zapping described in Sandhu, Pandey \& Sagar (2003).
For a randomly selected
star in the $V$, $(V - I)$ CMD of the field region, the nearest
star in the cluster's $V$, $(V - I)$ CMD within $V$ $\pm$ 0.35 and
$(V -I)\pm$0.2 of the field star was removed. While removing
stars from the cluster CMD, necessary corrections for 
incompleteness of the data samples were taken into account. 
The statistically cleaned $V, (V - I)$ CMD
of the cluster region is shown in Figure~\ref{fig:st_cmdvi} (c) which clearly shows the presence of 
PMS   
stars in the cluster.

\begin{figure*}
  \includegraphics[width=13.0cm]{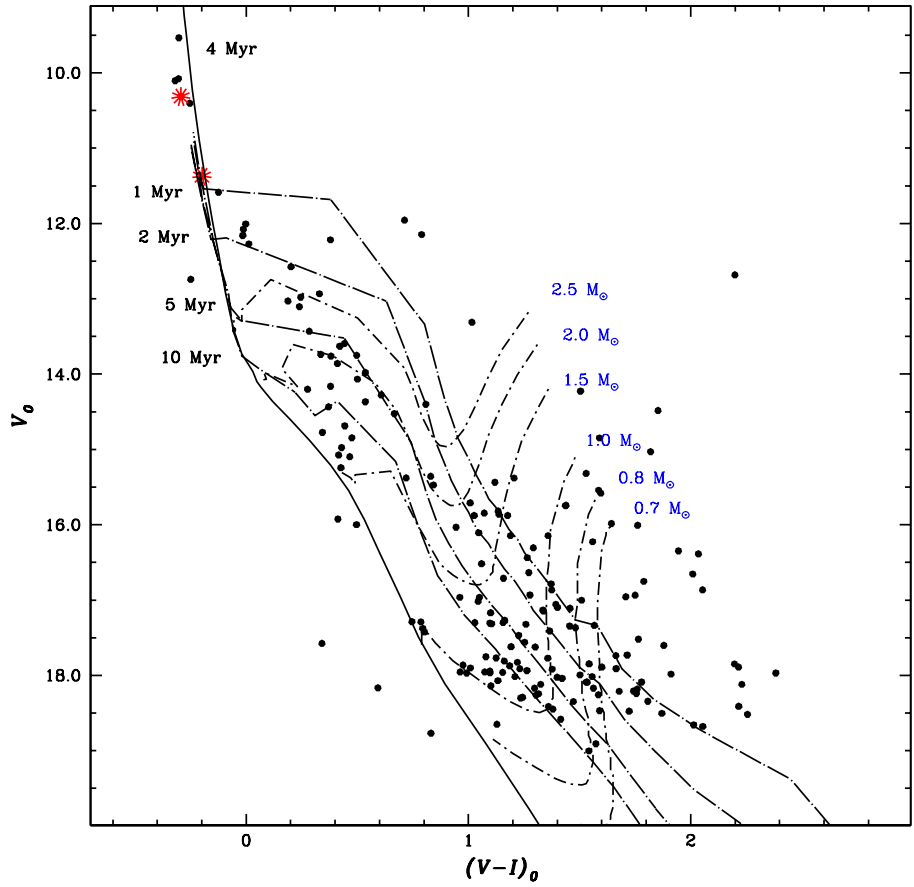}
  \caption{Statistically cleaned $V_0 /(V -I)_0$ CMD for stars lying
in the cluster region. The isochrone for 4 Myr age by Girardi et
al. (2002) and PMS    isochrones of 1,2,5,10 Myr along with 
evolutionary tracks of different mass stars by Siess, Dufour \& Forestini (2000) are
also shown. All the isochrones are corrected for a distance of 2.8
kpc. 
 Early B-type stars Sh 2-170 3 and LS I +64 9 are marked by 
the symbol of asterisk.
}
  \label{fig:st_cmd_sfr}
  \end{figure*}

Figure~\ref{fig:st_cmd_sfr} shows
statistically cleaned unreddened $V_0, (V - I)_0$ CMD.
The stars
having spectral type earlier than A0 were individually unreddened (cf. \S\ref{sec:st_ext}), whereas
mean reddening of the region, estimated from available individual reddening values in
that region, was used for other stars (cf. Sharma et al. 2007 ). 
The isochrone for 4 Myr
 by Girardi et al. (2002) and PMS    isochrones by 
Siess, Dufour \& Forestini (2000) have been plotted in Figure~\ref{fig:st_cmd_sfr}.
 Evolutionary
tracks by Siess, Dufour \& Forestini (2000) for various masses have also been shown in the same figure. 

\subsection{Mass function} \label{sec:st_mf}

\begin{figure*}
  \centering 
 \includegraphics[width=10cm]{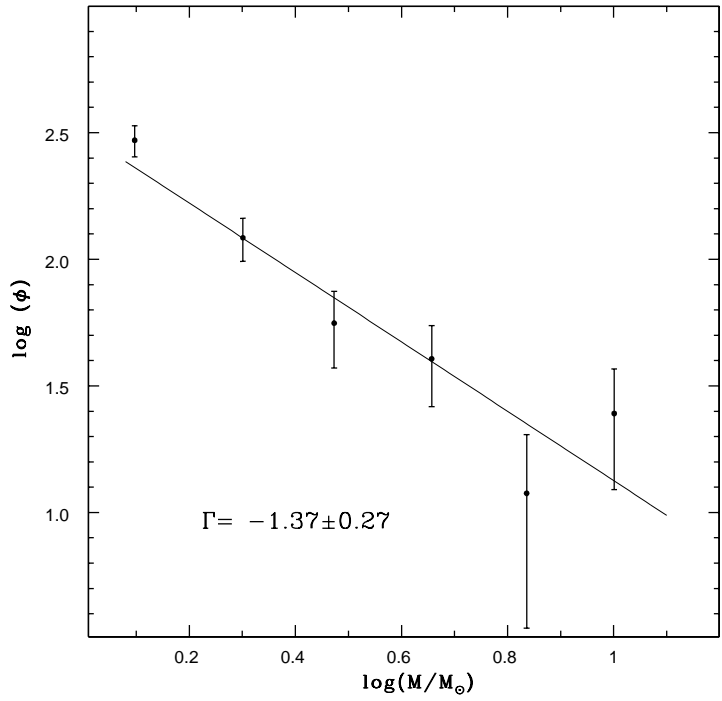} 
  \caption{A plot of the MF in the cluster. log($\phi$) 
represents log $(N /d (log~m))$. The error bars represent $\pm \sqrt{N}$ errors.
The solid line shows a least square fit to the entire mass range
1.00 $< M/M_\odot <$ 11.88.}
  \label{fig:st_mf}
  \end{figure*}

With the help of statistically cleaned CMD, shown in Figure~\ref{fig:st_cmd_sfr}, 
we can derive the mass function (MF)
using the theoretical evolutionary model of Girardi et al. (2002). 
Here, we are not discussing the mass function of the stars brighter than $V_0 <$ 9.5 mag ($V<$12.0 mag)  because
they are saturated even on our shortest exposure.  
Since 
post-main-sequence age of the cluster is  6 Myr, the stars having $V_0 <$ 13.5 ($M >$2.5$M_\odot$)
 have been considered to be on the MS.
For the MS stars, the LF was
converted to the MF using the theoretical model by Girardi et al. (2002)
 (cf. Pandey
et al. 2001, 2005). The MF for PMS stars was obtained by counting the number of
stars in various mass bins 
(shown as evolutionary tracks; Siess, Dufour \& Forestini (2000)) in Figure~\ref{fig:st_cmd_sfr}. 
The  MF  for the whole cluster is given in  Table~\ref{tab:st_mf} 
and plotted in  Figure~\ref{fig:st_mf}.
The slope of the MF for the entire observed mass range 1.0$<M/M_{\odot}<$11.9 
is derived to be $\Gamma$ = -1.37 $\pm$ 0.27, which is in agreement with the  
Salpeter value ( = -1.35; Salpeter 1955).

\subsection{Mass segregation} \label{sec:st_mass_seg}
\begin{figure*}
  \centering 
 \includegraphics[width=10cm]{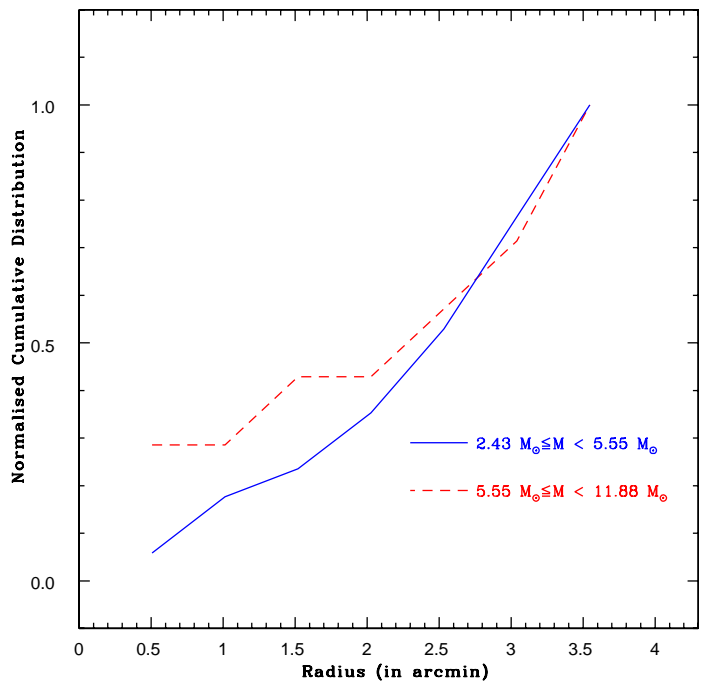} 
  \caption{ Cumulative radial distribution of MS stars in two
mass intervals.}
  \label{fig:st_ms}
  \end{figure*}

To characterize the degree of mass segregation in Stock 18, the MS
sample is subdivided into two mass groups (5.55 $\leq$ $M/M_\odot < $ 11.88, 2.43 $\leq$ $M/M_\odot < $ 5.55).
Figure~\ref{fig:st_ms}
shows cumulative distribution of MS stars as a function of radius in two different mass
groups. The figure indicates that more massive stars (5.55 $\leq$ $M/M_\odot < $ 11.88) tend to lie toward
the cluster center. However, the Kolmogorov-Smirnov test shows that the confidence level
with the statement is very low, i.e., 45\%.
Therefore, no statistically significant effect of mass 
segregation is found in the cumulative distribution.

\section{2MASS NIR  data}
\label{sec:st_nir}
 The dust grains in the circumstellar 
disk of PMS stars absorb a fraction of visible or ultraviolet  photons emitted by central star,
and re-radiates the absorbed energy in IR wavelengths. Thus, the presence of NIR excess 
can be used to distinguish  between the PMS stars from other stars in optical CMD.
Therefore, NIR observations are very effective for investigating the nature of obscured clusters
and the populations of YSOs which are embedded in molecular clouds (Lada \& Adams
1992).

\begin{figure*}
  \centering 
 \includegraphics[width=8cm]{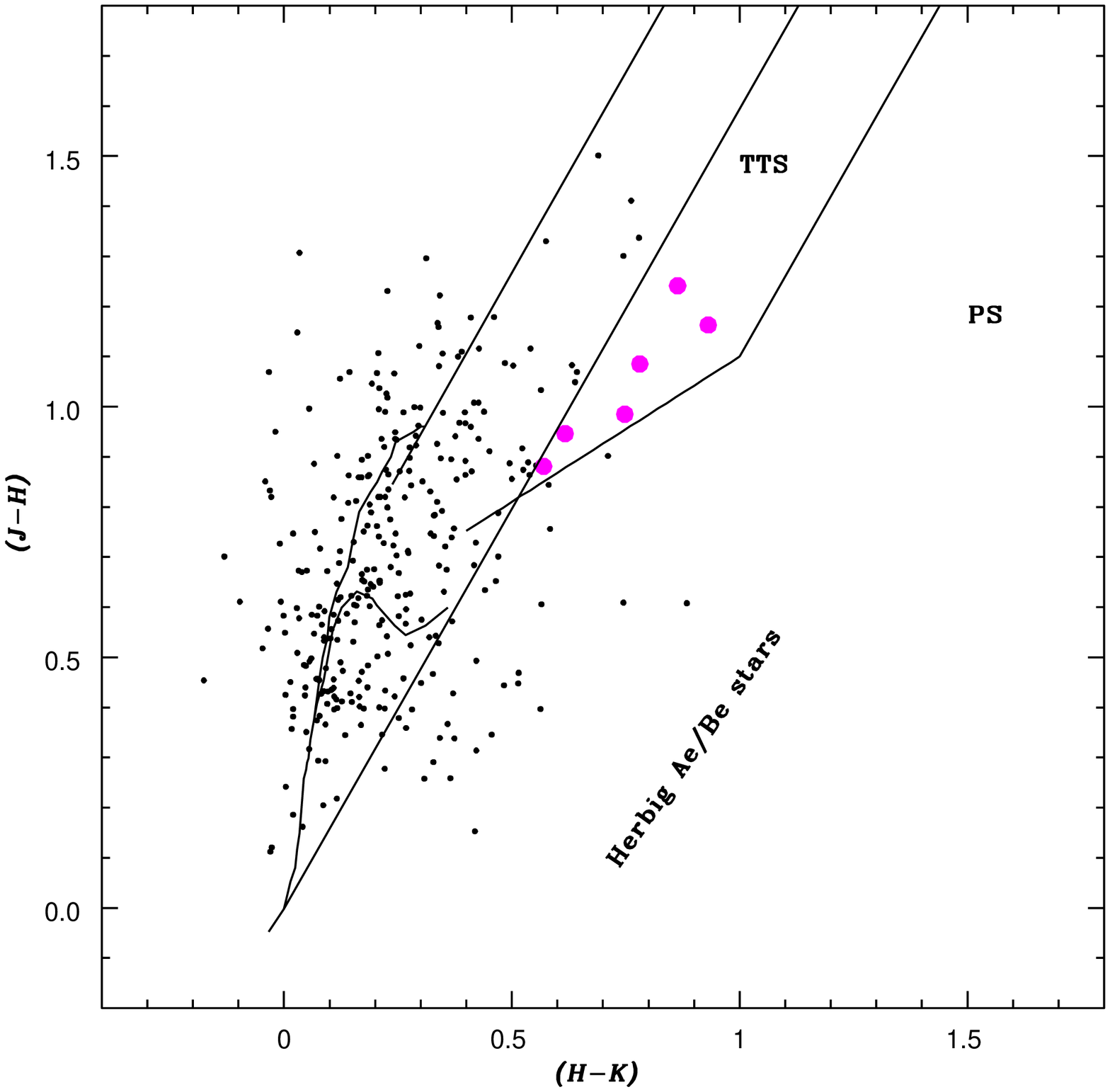} 
 \includegraphics[width=8cm]{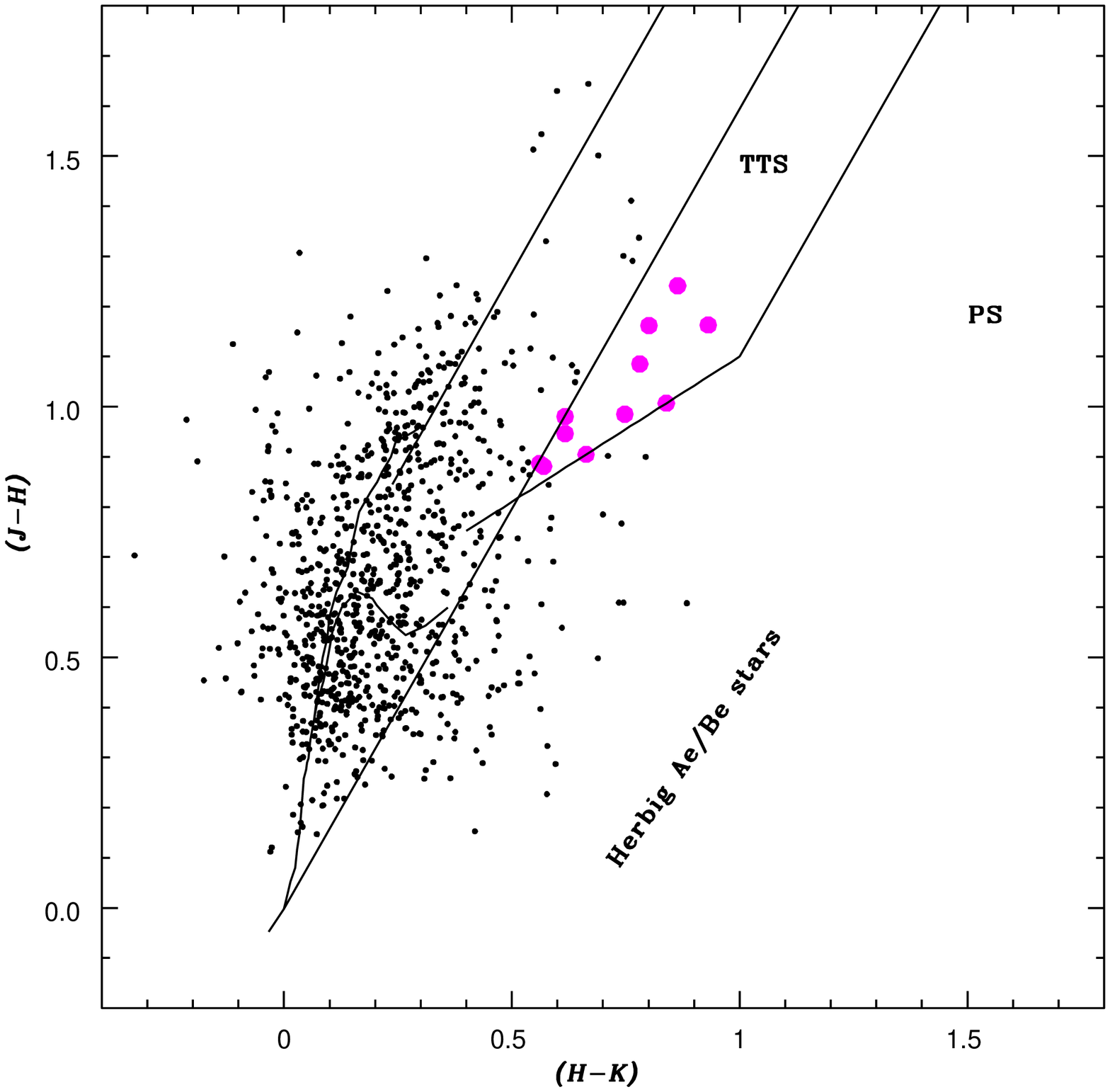}
  \caption{CC diagrams using the 2MASS JHK data. 
    {\it upper panel:} The NIR CC 
    diagram for the cluster region Stock 18.
    {\it lower panel :} Same as the upper panel, but for the 0.5 degree 
    surrounding region centered around the cluster and represent the Sharpless region Sh2-170.
     Filled circles in TTS region denote T-Tauri stars with NIR excess.}
  \label{fig:st_NIRccd}
  \end{figure*}

Figure~\ref{fig:st_NIRccd} shows $(J-H),(H-K)$ CC diagram for the cluster region ($r < r_{\rm{cl}}$) and 0.5 degree region centered around the cluster which represents the 
Sharpless region Sh-2 170.
Unreddened 
MS and giant branch, taken from Bessell \& Brett(1988), are represented by solid lines.
The reddening vectors for early and late-type stars  are shown by  parallel 
solid lines drawn from the base and the tip of two branches.
 Location of T-Tauri 
stars (marked by TTS; Meyer, Calvet \& Hillenbrand 1997), proto-star (marked by PS) like objects 
and Herbig Ae/Be (Hernandez et al. 2005) are 
also shown.
 The extinction ratio $A_J/A_V=0.282$, $A_H/A_V=0.180$
and $A_K/A_V=0.116$ have been taken from Cardelli, Clayton \& Mathis (1989). 
Stars below the reddening vectors in TTS region in Figure~\ref{fig:st_NIRccd} are considered to be NIR excess stars.
These stars are young with the characteristics of having  circumstellar material.
We found six such stars in the cluster region, however, the Sharpless 
region Sh-2 170 contains eleven  NIR excess stars. Therefore, nearly 54\% of 
total stars present in Sharpless region Sh-2 170
are found within cluster region ($r<r_{\rm{cl}}$). Carpenter (2000) estimated that 
55\% of the YSOs present in 
SFRs are
contained in the clusters. However, Bressert et al. (2010) showed that only a 
low fraction ($<$26\%) of YSOs are formed
in dense environments using Spitzer Space Telescope surveys.  Out of the six NIR excess stars in cluster region, only four stars have optical counterpart.

\begin{figure*}
\includegraphics[width=14.5cm]{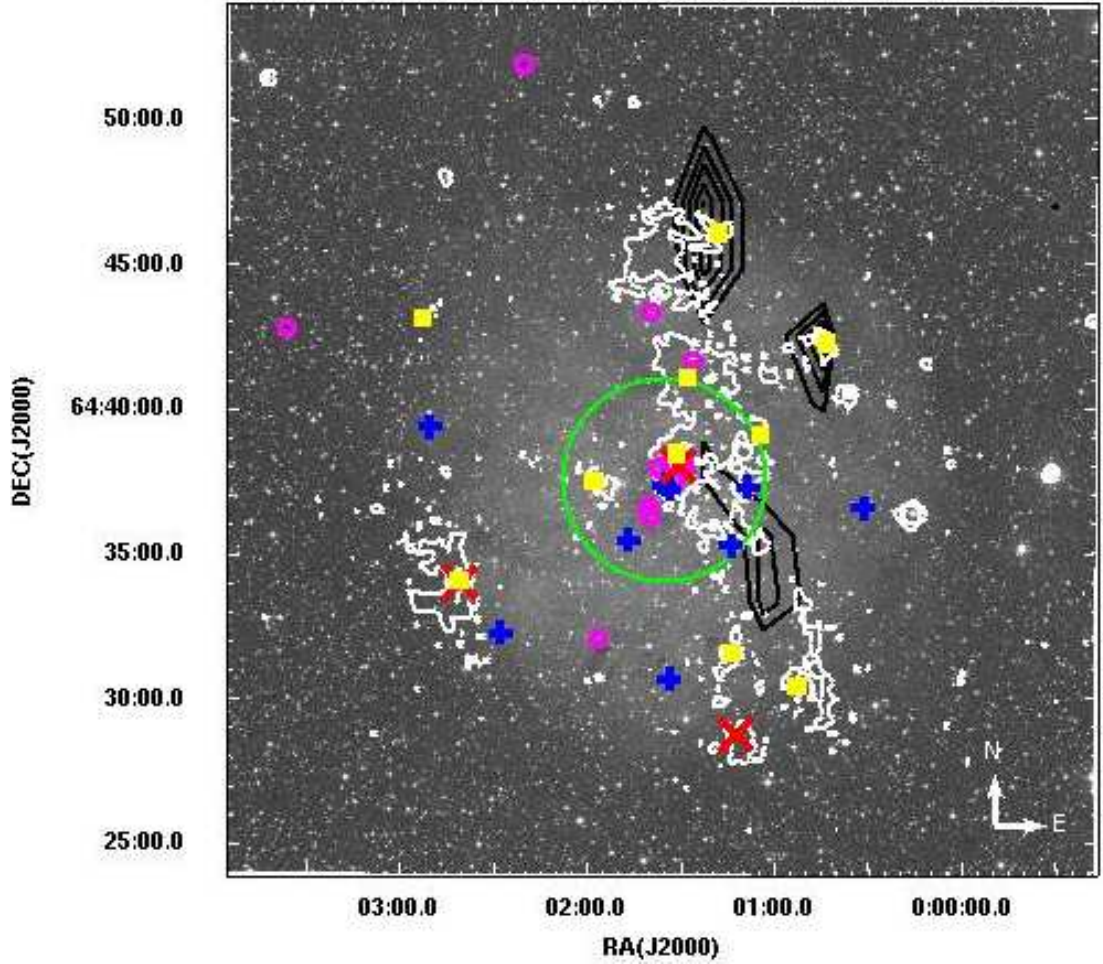} 
\caption{Spatial distributions of O/B-type stars (plus symbol), 
IR-excess sources (probable T-Tauri type stars, small open circles), 
IRAS points sources (crosses) and AKARI sources (open squares) are 
overlaid on the image of Sh2-170 in DSS-II R band. 
Cluster region Stock 18 ($r<r_{\rm{cl}}$) is marked by big open circle. 
${}^{12}$CO contours (black contours) from Kerton \& Brunt (2003)
and MSX A-Band intensity contours (white contours ) have also been shown. 
The MSX A-Band contours are at 0.12, 2.24, 2.41, 2.55, 2.68, 2.85, 5.26 ${\rm{\times 10^{-5}~W~m^{-2}~Sr^{-1}}}$.
${}^{12}$CO contours are at 9.0, 9.3, 9.6, 9.9, 10.2, 10.5 K.
}
\label{fig:st_dis_co_msx}
\end{figure*}

    In Figure~\ref{fig:st_dis_co_msx}, spatial distribution of O/B-type stars (plus symbol),  IR-excess sources (open circles), IRAS points sources (crosses) and AKARI sources (open squares) are 
displayed on a 30$^{\arcmin}$ $\times$ 30$^{\arcmin}$  DSS-II R band
image around the cluster. 
 The MSX A-band intensity map is superimposed on DSS-II R band image in Figure~\ref{fig:st_dis_co_msx},
which indicates presence of several discrete sources 
 representing high-density clumps and region of further star
formation. 
The $\rm{^{12}}$CO temperature map 
of the region has also been shown in the Figure~\ref{fig:st_dis_co_msx} using the data from Kerton \& Brunt (2003).
The AKARI  and IRAS sources are associated with MSX intensity map
as well as with $\rm{^{12}}$CO map, indicate the youth of these
regions.  
The presence of FIR IRAS and AKARI sources which may even younger than 1 Myr, supports the
ongoing star formation process within Sh-2 170 region.

\begin{figure*}
  \includegraphics[width=13.0cm]{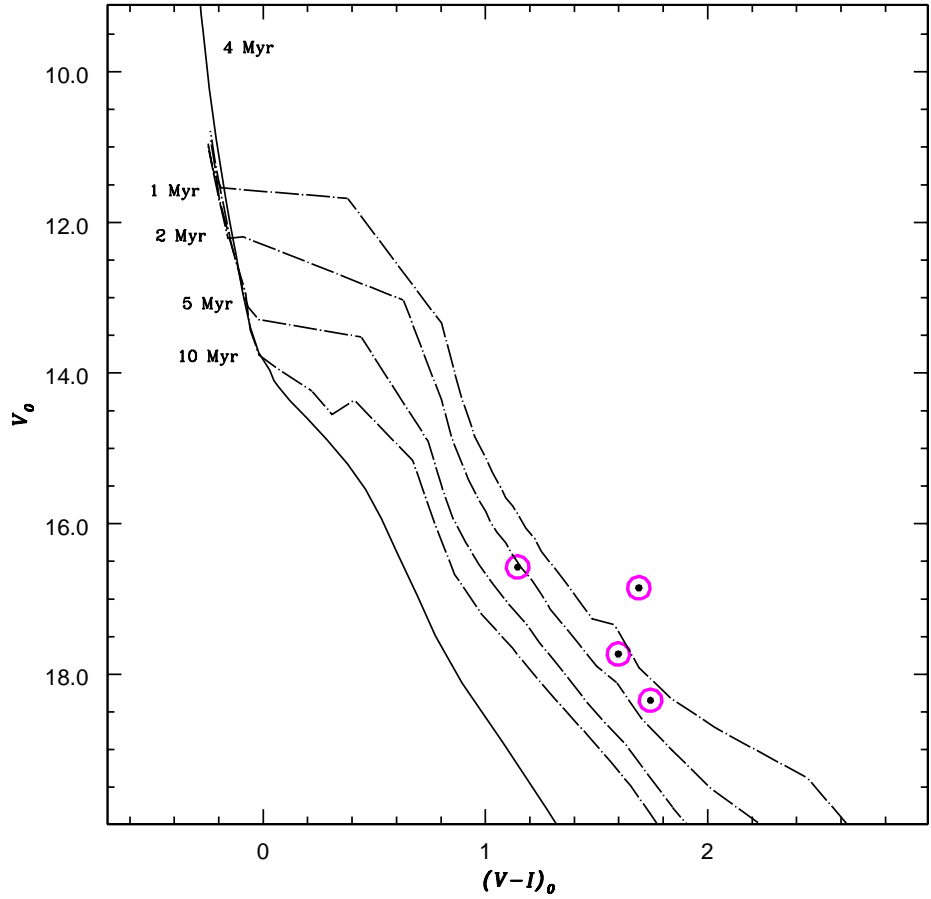}
  \caption{NIR excess sources (see \S\ref{sec:st_nir}) within the cluster region in $V_0 /(V -I)_0$ CMD.
 The isochrone for 4 Myr age by Girardi et
al. (2002) and PMS    isochrones of 1,2,5,10 Myr along with 
evolutionary tracks of different mass stars by Siess, Dufour \& Forestini (2000) are
also shown. All the isochrones are corrected for a distance of 2.8
kpc. 
}
  \label{fig:st_cmd_agespread}
  \end{figure*}

The locations of NIR excess sources within cluster region in optical $V_0, (V - I)_0$ CMD are shown in Figure~\ref{fig:st_cmd_agespread}.
The isochrone for 4 Myr
 by Girardi et al. (2002) and PMS   isochrones by 
Siess, Dufour \& Forestini (2000) are also  over plotted in Figure~\ref{fig:st_cmd_agespread}.
These probable YSOs are individually unreddened using NIR data.
The NIR CC
diagrams were used to estimate $A_V$ for each of these YSOs by tracing them back to the
intrinsic CTTS locus of Meyer et al. (1997) along the reddening vector (for details see Ogura et al. 2007). It appears that the age of these NIR excess PMS stars is 1-2 Myr. 
However, the estimation of the age of the PMS stars by comparing the observations
with the theoretical isochrones is prone to the random as well as systematic errors (see e.g., Hillenbrand 2005, Hillenbrand et al. 2008). 
The other sources of errors which can affect the estimation of ages may be the intrinsic variability of YSOs, accretion processes, uncertainties in determination of extinction, 
binarity and dispersion in distance measurement (Hartmann 2001). 
Binarity will brighten the star, consequently the CMD will yield a lower age estimate. In the case of equal mass binaries, we expect an error of  50 - 60\% in the age estimation of the PMS stars. However, it is difficult to estimate the influence of binarity on the mean age estimation as
the fraction of binaries is not known. If the PMS stars are cluster members, it appears that there is difference in age of YSOs and massive star BD +63 2093p. 
Moreover, the PMS stars are located in the clumps (please see the Figure 10) and star BD+63 2093 is located out side of these clumps within the cluster region. In past, it was also noticed that ongoing star formation may occur in the vicinity of massive star with in a very small angular diameter even 1 pc ( e.g. Azimlu \& Fich 2011). 
However, current data are not enough to make a confident claim on the ongoing star formation and more FIR
and deep optical data is needed to probe the real picture of star formation in the
region of the cluster Stock 18.

\section{Summary and Conclusion} \label{sec:st_summary}

On the basis of a comprehensive multi-wavelength study, here, 
an attempt has been made to understand the basic parameters and the implications of star formation processes
in the cluster Stock 18. 
Deep optical \ubvri  data along with archival data from the surveys such as 2MASS, MSX, IRAS, AKARI
and $\rm{^{12}}$CO are used to understand the global scenario of star formation in and around the
cluster region. 
    Reddening, $E(B - V )$, in the direction of cluster is found to be normal but varying between
0.7 to 0.9 mag. The post-main-sequence age and distance to the cluster are found to
be  6$\pm$2 Myr and 2.8$\pm$0.2 kpc respectively. 
The  slope of the MF
is found to be -1.37$\pm$0.27 for the mass range 1.0$<M/M_{\odot}<$11.9.
 The effect of mass segregation is not seen in the MS
stars.
A population of PMS YSOs in the cluster region having masses  0.8 - 2 $M_{\odot}$ have been found within the cluster.
 The position
of the YSOs on the CMDs indicates that  these stars have age between 
1 to 2 Myr.  However, this age spread is affected with observational errors and limitations of theoretical models. Further, the present data sets are not enough to explain the ongoing star formation.  


\section*{Acknowledgments}
Authors are thankful to anonymous referee for his/her useful comments which has
improved contents and presentation of the paper significantly.
We acknowledge the archival data sources  
the Two Micron All Sky Survey, which is a joint project of
the University of Massachusetts; the Infrared Processing and
Analysis Center/California Institute of Technology, funded
by the National Aeronautics and Space Administration and
the National Science Foundation and VizieR catalogue 
access tool, CDS, Strasbourg, France.  
This research makes use of observations with AKARI, a JAXA project with the participation of ESA.

\clearpage
\begin{table*}
\caption{Observation log of CCD observations of the cluster Stock 18 and the 
  calibration region SA98 (Landolt 1992).}             
\label{tab:st18_optlog}      
\centering                          
\begin{tabular}{l c l l }        
\hline\hline                 
           Date(UT)& Filter & \multicolumn{2}{c}{Exposure Time (s)}                  \\      
                   &        & \multicolumn{2}{c}{($\times$no. of exposures)}         \\  
\hline                                                      
                   &                   &  Stock 18                        & SA 98 \\
\hline                                                              
   08 October 2007 & $U$               & 1800$\times$2              &   \\
                   & $B$               & 1200$\times$3              &   \\
                   & $V$               & 900$\times$3               &   \\
                   & $R$               & 300$\times$3               &   \\
                   & $I$               & 300$\times$3               &    \\
   19 December 2006& $V$               & 900$\times$3               &\\
                   & $R$               & 480$\times$2,60$\times$1   &\\
                   & $I$               & 480$\times$2,300$\times$2  &\\
  23 December 2006 & $U$               & 300$\times$2               & 300$\times$7\\
                   & $B$               & 120$\times$3               & 120$\times$7\\
                   & $V$               & 120$\times$3               & 60$\times$7\\ 
                   & $R$               & 40$\times$3                & 40$\times$6\\
                   & $I$               & 50$\times$5                & 40$\times$7\\
\hline                                                      
   \end{tabular}
   \end{table*}

\begin{table}
\tiny
\centering
\caption{\normalsize $UBVRI$ photometric data of the sample stars in 13$\arcmin$$\times$13$\arcmin$ region.
CCD positions of stars  are converted into $\rm{RA_{J2000}}$ and 
$\rm{DEC_{J2000}}$ using the Guide Star Catalogue II (GSC 2.2, 2001). The complete table is
available in electronic form.
}
\label{tab:datasample}
\begin{tabular} {c c c c c c c c c c }
\hline
ID & $\rm{RA_{J2000}}$  & $\rm{DEC_{J2000}}$ &  $U$ &$B$&$V$&$R$&$I$\\
   &(deg) & (deg) & (mag) &(mag)&(mag)&(mag)&(mag)\\
\hline

1     &  0.388444  &   64.512169  & 11.539 $\pm$   0.008 &   12.016   $\pm$  0.002 & 11.608  $\pm$   0.008 &   11.386  $\pm$   0.002 & 11.089   $\pm$  0.002\\ 
    2     &  0.381583  &   64.677193  & 12.609 $\pm$   0.008 &   12.878   $\pm$  0.003 & 12.241  $\pm$   0.004 &   11.882  $\pm$   0.007 & 11.453   $\pm$  0.005\\ 
    3     &  0.417500  &   64.531441  & 14.559 $\pm$   0.038 &   13.639   $\pm$  0.005 & 12.247  $\pm$   0.004 &   11.436  $\pm$   0.026 & 10.595   $\pm$  0.031\\ 
    4     &  0.306694  &   64.588333  & 12.618 $\pm$   0.009 &   13.007   $\pm$  0.004 & 12.542  $\pm$   0.003 &   12.274  $\pm$   0.003 & 11.938   $\pm$  0.003\\ 
--&--&--&--&--&--& --&--\\ 
\hline
\end{tabular}
\end{table}

\begin{table*}
\caption{The MF of the cluster Stock 18.  The numbers of probable cluster members (N) have been
 obtained after subtracting the expected contribution of field stars in \S\ref{sec:st_cmd}. log$\phi$ 
represents log $(N/d (log~m))$.}             
\label{tab:st_mf}      
\centering                          
\begin{tabular}{c c c c c c c c c}       
\hline                 
Magnitude range     & Mass  Range    & Mean            &  \multicolumn{2}{c} { Whole region}\\
    ($V_0$ mag)     &   (M$_\odot$)  &  log (M$_\odot$)&    N     & log$\phi$               \\
\hline
&&&&\\
\hline
 9.5 - 10.5 &  11.88 -  8.17 &  1.001 &   4 &  1.391\\
10.5 - 11.5 &   8.17 -  5.55 &  0.836 &   2 &  1.076\\
11.5 - 12.5 &   5.55 -  3.52 &  0.657 &   8 &  1.607\\
12.5 - 13.5 &   3.52 -  2.43 &  0.473 &   9 &  1.748\\
            &   2.50 -  1.50 &  0.301 &  27 &  2.085\\
            &   1.50 -  1.00 &  0.097 &  52 &  2.470\\
\hline
\end{tabular}
\end{table*}

\clearpage

\newpage

\clearpage

\begin{thebibliography}{}
\bibitem[Azimlu et al. (2011)]{azimlu2011}Azimlu M., Fich M., 2011, \aj, 141, 123
\bibitem[Bessell \& Brett (1988)]{bessell88} Bessell M. S., Brett J. M., 1988, \pasp, 100, 1134
\bibitem[Bressert et al. (2010)]{bressert2010}Bressert E., Bastian N., Gutermuth R., Megeath S. T., Allen L., Evans N. J. II, Rebull L. M., Hatchell J. et al., 2010, \mnras, 409, 54 
\bibitem[Caldwell et al.(1993)]{caldwell93}Caldwell A. R. John, Cousins A. W. J., Ahlers C. C., Wamelen P. van, Maritz E. J., 1993, SssAAO Circ. 15
\bibitem[Cardelli, Clayton \& Mathis(1989)]{cardelli89} Cardelli J. A., Clayton G. C., Mathis J. S., 1989, \apj, 345, 245
\bibitem{Carpenter}Carpenter J. M., 2000, AJ, 120, 3139
\bibitem{Chini}Chini R., Wargau W. F., 1990, A\&A, 227, 213
\bibitem[Girardi et al.(2002)]{girardi02}  Girardi L., Bertelli G., Bressan A., Chiosi C., Groenewegen M. A. T.,  Marigo P., Salasnich B., Weiss A., 2002, \aa, 391, 195
\bibitem{Hartmann2001}Hartmann L., 2001, AJ, 121, 1030
\bibitem{Hillenbrabd2005}Hillenbrand L.A., 2005, A Decade of Discovery: Planets Around Other Stars" STScI 
Symposium Series 19, ed. M. Livio, eprint arXiv:astro-ph/0511083
\bibitem{Hillenbrabd2008}Hillenbrand L.A., Bauermeister A., White R.J., 2008, in ASP Conf. Ser. 384, 14th 
Cambridge Workshop on Cool Stars, Stellar Systems, and the Sun, ed. G. van Belle (San Francisco
CA: ASP), 200
\bibitem[Hernandez et al.(2005)]{hernandez05}  Hernandez J., Calvet N., Hartmann L., Briceno C., Sicilia-Aguilar A., Berlind P., 2005, \aj, 129, 856
\bibitem[Johnson \& Morgan(1953)]{johnson53} Johnson H. L., Morgan W. W., 1953, ApJ, 117, 313
\bibitem[Johnson(1968)]{johnson68}Johnson H. L., 1968, in Middlehurst B. M., Aller, L.H., eds,  Nebulae and Interstellar Matter. Univ. Chicago, 191
\bibitem{joshih}Joshi H., Kumar B., Singh K. P., Sagar R., Sharma S., Pandey J. C., 2008, \mnras, 391, 1279 
\bibitem[Kerton \& Brunt(2003)]{kerton03}Kerton C. R., Brunt C. M., 2003, \aa, 399, 1083
\bibitem[King(1962)]{king62} King I., 1962, \aj, 67, 471
\bibitem[Koornneef(1983)]{koornneef83} Koornneef J., 1983, \aa, 128, 84
\bibitem{Lada92}Lada C. J., Adams F. C., 1992, AJ, 393, 278
\bibitem[Landolt(1992)]{landolt92}  Landolt A. U., 1992, \aj, 104, 340 
\bibitem{mayer73}Mayer P., Mac$\rm{\acute{a}}$k P., 1973, Bull. Astron. Inst. Czech., 24, 50
\bibitem[Meyer, Calvet \& Hillenbrand(1997)]{meyer97} Meyer M., Calvet N., Hillenbrand L. A., 1997, \aj, 114, 288
\bibitem{Ogura07}Ogura K., Chauhan N., Pandey A.K., Bhatt B.C., Ojha D.K., Itoh Y., 2007, PASJ, 59, 199
\bibitem{Pandey00}Pandey A. K., Ogura K., Sekiguchi K., 2000, PASJ, 52, 847
\bibitem[Pandey et al.(2001)]{pandey01}Pandey A. K., Nilakshi, Ougra K., Sagar R., Tarusawa K., 2001, \aa, 374, 504
\bibitem[Pandey et al.(2005)]{pandey05} Pandey A. K., Upadhyay K., Ougra K., Sagar R., Mohan V., Mito H., Bhatt H. C., Bhatt B. C., 2005, \mnras, 358, 1290
\bibitem{Roger}Roger R. S., McCutcheon W. H., Purton c. R., Dewdney P. E., 2004, A\&A, 425, 553
\bibitem{Russeil}Russeil D., Adami C. , Georgelin Y. M., 2007, A\&A, 470, 161 
\bibitem[Sagar \& Richtler(1991)]{sagar91}  Sagar R., Richtler T., 1991, \aa, 250, 324
\bibitem[Salpeter(1955)]{salpeter55} Salpeter E. E., 1955, \apj, 121, 161
\bibitem[Sandhu, Pandey \& Sagar(2003)]{sandhu03}  Sandhu T. S., Pandey A. K., Sagar R, 2003, \aa, 408, 515
\bibitem[Schmidt-Kaler(1982)]{schmidtkaler82} Schmidt - Kaler Th., 1982, In: Landolt/Bornstein, Numerical Data and  Functional Relationship in Science and Technology, New series, Group VI, Vol.  2b, Scaifers K. \& Voigt H. H. (eds.) Springer - Verlog, Berlin, p. 14
\bibitem[Sharma et al. (2007)]{sharma07} Sharma S., Pandey A. K., Ojha D. K., Chen W. P., Ghosh S. K., Bhatt B. C., Maheswar G., Sagar R., 2007, MNRAS, 380, 1141
\bibitem[Siess, Dufour \& Forestini(2000)]{siess00} Siess L., Dufour E., Forestini M., 2000, \aa, 358, 593
\bibitem{Skiff09}Skiff B. A., 2009, yCat, 1, 1023S, Available at VizieR On-line Data Catalog. http://webviz.u-strasbg.fr/viz-bin/VizieR
\bibitem[Stetson(1987)]{stetson87}  Stetson P. B., 1987, \pasp, 99, 191
\bibitem[Stetson(1992)]{stetson92} Stetson P. B., 1992, in  ASP Conf. Ser. Vol. 25, Astronomical Data Analysis  Software and Systems I.  Astron. Soc. Pac., ed.  Worrall D. M., Biemesderfer C., Barnes J., San Francisco, 297
\bibitem[Whittet \& van Breda(1980)]{whittet80}  Whittet D. C. B., van Breda I. G., 1980, \mnras, 192, 467
\bibitem[Zavagno et al. (2010)]{Zavagno2010}Zavagno A., Russeil D., Motte F., Anderson L. D., Deharveng L., Rod$\acute{o}$n J. A., Bontemps S., Aberge A. et al., 2010, \aa, 518, L81 
\end{thebibliography}
\end{document}